\documentclass[conference]{IEEEtran}
\IEEEoverridecommandlockouts
\usepackage{cite}

\usepackage{soul}
\usepackage{amsmath,amssymb,amsfonts}
\usepackage{algorithmic}
\usepackage{graphicx}
\usepackage{textcomp}
\usepackage{xcolor}
\usepackage{textgreek}
\usepackage{float}
\usepackage{siunitx}
\usepackage{hyperref}

\usepackage{braket}

\definecolor{indigo}{rgb}{0.29, 0.0, 0.51}
\definecolor{neongreen}{rgb}{0.22, 0.88, 0.08}

\def\digamma{\mbox{\DejaVuSans\char"03DD}}

\def\BibTeX{{\rm B\kern-.05em{\sc i\kern-.025em b}\kern-.08emhttps://www.overleaf.com/project/612e1ceea68183218f0e46a4
    T\kern-.1667em\lower.7ex\hbox{E}\kern-.125emX}}
\begin{document}

\title{Bimodal Approach for Noise Figures of Merit Evaluation in Quantum-Limited Josephson Traveling Wave Parametric Amplifiers

{\footnotesize}
\thanks{This work is supported by the Italian Institute of Nuclear Physics (INFN) within the Technological and Interdisciplinary research commission (CSN5), by the European Union’s H2020-MSCA Grant Agreement No.101027746, by the Institute for Basic Science (IBS-R017-D1) of the Republic of Korea, by the University of Salerno - Italy under the projects FRB19PAGAN and FRB20BARON, by the SUPERGALAX project in the framework of the H2020-FETOPEN-2018-2020 call, and the Joint Research Project PARAWAVE of the European Metrology Programme for Innovation and Research (EMPIR). This project (PARAWAVE) received funding from the EMPIR programme co-financed by the Participating States and from the European Union Horizon 2020 research and innovation programme.\\
\\
(\textit{Corresponding author: Emanuele Enrico. Email:e.enrico@inrim.it})\\
L. Fasolo and A.Greco are with INRiM, Istituto Nazionale di Ricerca Metrologica, I-10135 Torino, Italy and with Politecnico di Torino, I-10129 Torino, Italy (luca\_fasolo@polito.it, a.greco@inrim.it).

E. Enrico and L.Oberto are with INRiM, Istituto Nazionale di Ricerca Metrologica, I-10135 Torino, Italy and with INFN - Trento Institute for Fundamental Physics and Applications, I-38123, Povo, Trento, Italy (e.enrico@inrim.it, l.oberto@inrim.it).

C. Barone, G. Carapella, and S. Pagano are with University of Salerno, Department of Physics, I-84084 Fisciano, Salerno, Italy, with INFN - Napoli, Salerno group, I-84084 Fisciano, Salerno, Italy, and with CNR-SPIN Salerno section, I-84084 Fisciano, Salerno, Italy (cbarone@unisa.it, gcarapella@unisa.it, spagano@unisa.it).

M. Borghesi, M. Faverzani, E. Ferri, A. Giachero, and A. Nucciotti are with University of Milano Bicocca, Department of Physics, I-20126 Milano, Italy and with INFN - Milano Bicocca, I-20126 Milano, Italy (matteo.borghesi@unimib.it, marco.faverzani@unimib.it, elena.ferri@unimib.it, andrea.giachero@unimib.it, angelo.nucciotti@unimib.it).

A. P. Caricato, A. Leo, G. Maruccio, A. G. Monteduro, and S. Rizzato are with University of Salento, Department of Physics, I-73100 Lecce, Italy and with INFN - Lecce, I-73100 Lecce, Italy (annapaola.caricato@unisalento.it, angelo.leo@unisalento.it, giuseppe.maruccio@unisalento.it, annagrazia.monteduro@unisalento.it, silvia.rizzato@unisalento.it).

I. Carusotto is with INO-CNR BEC Center, I-38123 Povo, Trento, Italy, with University of Trento, Department of Physics, I-38123, Povo, Trento, Italy and with Fondazione Bruno Kessler, I-38123, Povo, Trento, Italy (iacopo.carusotto@ino.cnr.it).

W. Chung, Ç. Kutlu, A. Matlashov, Y. K. Sermertzidis, and S. Uchaikin are with Center for Axion and Precision Physics Research, Institute for Basic Science (IBS), KR-34051, Daejeon, Republic of Korea. Ç. Kutlu and Y. K. Sermertzidis are also with Korea Advanced Institute of Science and Technology (KAIST), Department of Physics, KR-34051, Daejeon, Republic of Korea (gnuhcw@ibs.re.kr, caglar@kaist.ac.kr, andrei@ibs.re.kr, yannis@kaist.ac.kr, uchaikin@ibs.re.kr).

A. Cian, P. Falferi, D. Giubertoni, B. Margesin, and A. Vinate are with Fondazione Bruno Kessler, I-38123, Povo, Trento, Italy and with INFN - Trento Institute for Fundamental Physics and Applications, I-38123, Povo, Trento, Italy. P. Falferi and A. Vinante are also with IFN-CNR, I-38123 Povo, Trento, Italy (acian@fbk.eu, giuberto@fbk.eu, margesin@fbk.eu, paolo.falferi@unitn.it, andrea.vinante@ifn.cnr.it).

D. Di Gioacchino, C. Gatti, C. Ligi, G. Maccarrone, L. Piersanti, and A. Rettaroli are with INFN - Laboratori Nazionali di Frascati, I-00044, Frascati, Rome, Italy. A. Rettaroli is also with University of Roma Tre, Departement of Physics, I-00146, Roma, Italy. (daniele.digioacchino@lnf.infn.it, claudio.gatti@lnf.infn.it, carlo.ligi@lnf.infn.it, giovanni.maccarone@lnf.infn.it, luca.piersanti@lnf.infn.it, alessio.rettaroli@uniroma3.it).

G. Filatrella is with University of Sannio, Department of Science and Technology, I-82100, Benevento and with INFN - Napoli, Salerno group, I-84084 Fisciano, Salerno, Italy (filatr@unisannio.it).

P. Livreri is with University of Palermo, Department of Engineering, I-90128, Palermo, Italy and with CNIT - Interuniversity Consortium for Telecommunication, National Laboratory for Radar and Surveillance Systems, I-56124, Pisa, Italy (patrizia.livreri@unipa.it). 

C. Mauro is with INFN - Napoli, Salerno group, I-84084 Fisciano, Salerno, Italy (cmauro@na.infn.it).

R. Mezzena is with University of Trento, Department of Physics, I-38123, Povo, Trento, Italy and with INFN - Trento Institute for Fundamental Physics and Applications, I-38123, Povo, Trento, Italy (renato.mezzena@unitn.it).

V. Pierro is with University of Sannio, Department of Engineering, I-82100 Benevento, Italy and with INFN - Napoli, Salerno group, I-84084 Fisciano, Salerno, Italy (pierro@unisannio.it).

M. Rajteri is with INRiM - Istituto Nazionale di Ricerca Metrologica, I-10135 Torino, Italy and with INFN - Torino, I-10125 Torino, Italy (m.rajteri@inrim.it).

}

}

\author{L. Fasolo, C. Barone, M. Borghesi, G. Carapella, A. P. Caricato, I. Carusotto, W. Chung, A. Cian,\\D. Di Gioacchino, E. Enrico, P. Falferi, M. Faverzani, E. Ferri, G. Filatrella, C. Gatti, A. Giachero,\\D. Giubertoni, A. Greco, Ç. Kutlu, A. Leo, C. Ligi, P. Livreri, G. Maccarrone, B. Margesin, G. Maruccio,\\
A. Matlashov, C. Mauro, R. Mezzena, A. G. Monteduro, A. Nucciotti, L. Oberto, S. Pagano, V. Pierro,\\L. Piersanti, M. Rajteri, A. Rettaroli, S. Rizzato, Y. K. Semertzidis, S. Uchaikin, and A. Vinante}
\maketitle
\begin{abstract}
The advent of ultra-low noise microwave amplifiers revolutionized several research fields demanding quantum-limited technologies.
Exploiting a theoretical bimodal description of a linear phase-preserving amplifier, in this contribution we analyze some of the intrinsic properties of a model architecture (i.e., an rf-SQUID based Josephson Traveling Wave Parametric Amplifier) in terms of amplification and noise generation for key case study input states (Fock and coherents). Furthermore, we present an analysis of the output signals generated by the parametric amplification mechanism when thermal noise fluctuations feed the device.\newline

\end{abstract}
\section{Introduction}
Nowadays, the technological progress in several fields of research, spanning from quantum computation and communication \cite{Nielsen2010,Krantz2019, Blais2020}, radio-astronomy \cite{Smith2013}, radio  detection  and  ranging \cite{Lloyd2008, Tan2008, Barzanjeh2015, Fasolo2021}, up to fundamental physics experiments \cite{Caldwell2017, Jeong2020, Backes2021}, led to the demand for ultra-low noise microwave amplifiers for broadband   high-fidelity readout. All these applications are deeply affected by the noise performances of such amplifiers. In this contribution, the standard Haus-Caves description of noise generation in an ideal bosonic phase-preserving linear amplifier is taken into account \cite{Haus1962, Caves1982, Caves2012}. In this representation the amplifier is considered as a two-ports black-box driven at a pump frequency $\omega_\text{p}$ that amplifies a bosonic input mode at frequency $\omega$. The amplification is associated with the creation of a second mode at frequency $\omega'=\omega_{\text{p}}-\omega$ (the so-called idler mode of a three-wave mixing parametric amplification \cite{Boyd2008}) that is commonly considered as an internal mode of the amplifier that causes the onset of noise at the output port. Here, we extend and give a different perspective of this description considering the case in which an uncorrelated idler mode is already present at the input port (i.e., considering a bimodal input field), analyzing the effect of the interaction between these modes inside the amplifier in terms of typical noise estimators.This operative condition may arise in real measurement setups where the amplifier is exploited, for instance, for the multiplexed readout of broadband signals \cite{Renger2021} or for the joint detection and amplification of probing signals in a microwave quantum illumination experiment \cite{Fasolo2021}.
\newline
The theoretical framework presented in this manuscript is supported with numerical simulations of the noise estimators for a realistic implementation of a quantum-limited amplifier\cite{Caves1982, Caves2012} such as the rf-SQUID based Josephson Travelling Wave Parametric Amplifier (JTWPA) \cite{Macklin2015, Zorin2016, Zorin2017, Fasolo2019, Greco2021}, which represents a promising realization of a microwave amplifier with high gain, large bandwidth and quantum-limited added noise \cite{Clerk2010}. The core of this device is a repetition of rf-SQUIDs, embedded in a coplanar waveguide, that generate a non-dissipative and highly non-linear superconducting metamaterial.
\section{Gain, Quantum Efficiency, and Noise Figure}
\begin{figure}[t]
    \centering
    \includegraphics[width=\linewidth]{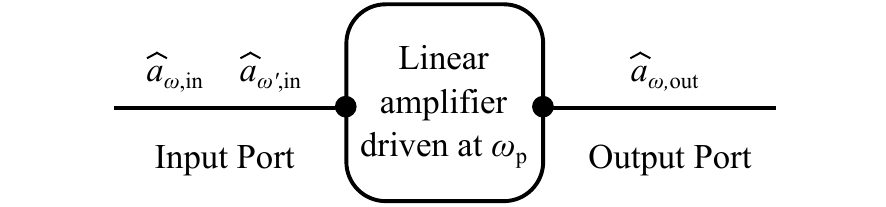}
    \caption{Pictorial representation of a linear amplifier, driven at $\omega_\text{p}$, having at the input port two uncorrelated fields at frequency $\omega$ and $\omega'=\omega_\text{p}-\omega$.}
    \label{fig:Schematic}
\end{figure}

Within the Heisenberg picture, a generic two-ports phase-preserving linear amplifier driven at frequency $\omega_{\text{p}}$ combines two uncorrelated input modes at frequency $\omega$ and $\omega'=\omega_\text{p}-\omega\neq\omega$, described through their dimensionless complex-amplitude operator $\hat{a}_{\omega,\text{in}}$ and $\hat{a}_{\omega',\text{in}}$, in an output mode at frequency $\omega$
\begin{equation}
\label{AnOutputOperator}
    \hat{a}_{\omega,\text{out}}=u(\omega)\hat{a}_{\omega,\text{in}}+iv(\omega)\hat{a}^{\dagger}_{\omega',\text{in}}
\end{equation}
with $[\hat{a}_{\omega,\text{in}},\hat{a}^\dagger_{\omega',\text{in}}]=\delta_{\omega,\omega'}$. A pictorial representation of the system is given in Fig.\;\ref{fig:Schematic}.
The output mode operator fulfils the common bosonic commutation relation only if the complex functions $u(\omega)$ and $v(\omega)$ respect the unitary condition 
\begin{equation}
\label{eq:UnitaryCondiction}
    |u(\omega)|^2-|v(\omega)|^2=1
\end{equation}
As reported in \cite{Greco2021} for an rf-SQUID based JTWPA, an expression of these functions can be analytically derived exploiting a circuit quantum electrodynamics description of the system. These quantities depend both on the design parameters of the device and on the driving conditions (i.e., on the pump frequency $\omega_\text{p}$ and on the amplitude of this tone, that can be quantified in terms of an intensity of current $I_\text{p}$). In this derivation, the ideality condition of the phase-preserving linear amplifier is guaranteed neglecting all the possible dissipation sources, such as the dielectric losses or the side-band generation \cite{Esposito2021}. All the numerical simulations given below are evaluated for the circuital parameters described in \cite{Greco2021} for an amplifier operating in the three-wave mixing regime and pumped with a current intensity $I_\text{p}$ at frequency $\omega_\text{p}=\SI{12}{\giga\hertz}$. 

The expectation value for the photon number of the output field is given by
\begin{align}
\label{ExptOutNumber}
    &\braket{\hat{n}_{\omega,\text{out}}}=\braket{\hat{a}^\dagger_{\omega,\text{out}}\hat{a}_{\omega,\text{out}}}=\nonumber\\
    &=|u(\omega)|^2\braket{\hat{n}_{\omega,\text{in}}}+|v(\omega)|^2\braket{\hat{n}_{\omega',\text{in}}}+\nonumber\\
    &+i\left(u^{*}(\omega)v(\omega)\braket{\hat{a}^\dagger_{\omega,\text{in}}\hat{a}^\dagger_{\omega',\text{in}}}-u(\omega)v^{*}(\omega)\braket{\hat{a}_{\omega',\text{in}}\hat{a}_{\omega,\text{in}}}\right)+\nonumber\\
    &+|v(\omega)|^2
\end{align}
where the first term is the input field at $\omega$ frequency being amplified of a factor $|u(\omega)|^2$, while the second term represents the contribution to the amplification of the input field at $\omega$ frequency given by the input field at $\omega'$. The third and fourth terms represent respectively the contributions deriving from the spontaneous annihilation or creation of a pump photon, that respectively implies the creation or the annihilation of a couple of photons at $\omega$ and $\omega'$. Eventually the last term, independent from the input fields, represents the number of photons with frequency $\omega$ added by the amplifier (the so-called added noise photons).

Supposing a vacuum state $\ket{\text{vac}}=\ket{0}_{\omega}\ket{0}_{\omega'}$ as the input state of the amplifier, one can easily derive from \eqref{ExptOutNumber} that the expectation value for the photon number of the output field is 
\begin{equation}
\label{eq:vacuum}
    \braket{\hat{n}^{\text{vac}}_{\omega,\text{out}}}=\bra{\text{vac}}\hat{n}_{\omega,\text{out}}\ket{\text{vac}}=|v(\omega)|^2
\end{equation}
This means that the added noise photons can be equivalently seen as the product of the amplification of an input vacuum state. Furthermore, from \eqref{ExptOutNumber} derives that the spectral gain distribution of the amplifier $G(\omega)$, meant as the scale factor of the sole input mode at frequency $\omega$, is given by
\begin{equation}
\label{eq:gain}
    G(\omega)\equiv
    |u(\omega)|^2
\end{equation}
whereas, in order to take into consideration the contribution given by the input field at $\omega'$ frequency, it is possible to define a bimodal gain $G_\text{b}(\omega)$ as
\begin{equation}
    G_\text{b}(\omega)\equiv\frac{\braket{\hat{n}_{\omega,\text{out}}}-\braket{\hat{n}^{\text{vac}}_{\omega,\text{out}}}}{\braket{\hat{n}_{\omega,\text{in}}}}
\end{equation}
This latter quantity reduces to $G(\omega)$ when the $\omega'$ input field is in the vacuum state.
\begin{figure}
    \centering
    \includegraphics[width=85mm]{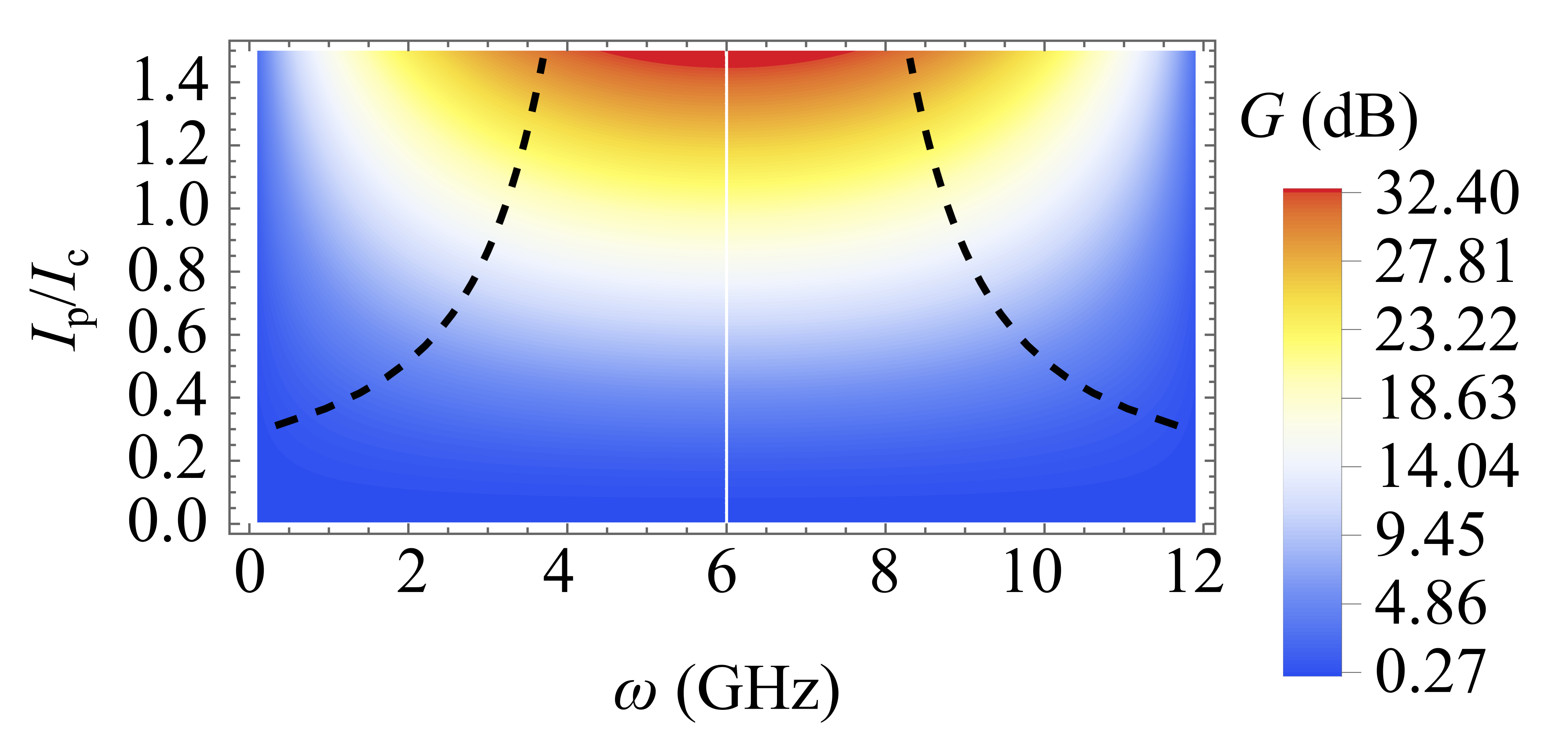}
    \caption{Gain spectrum $(G(\omega))$, expressed in dB,  as a function of $I_\text{p}$ normalized on the critical current $I_\text{c}$ of the Josephson junctions composing the referenced device. The dashed lines identify the bandwidth of this device at the 3dB threshold level}.
    \label{fig:Gain}
\end{figure}

Fig.\;\ref{fig:Gain} reports the spectral distribution of the gain $G(\omega)$ as a function of $I_\text{p}$ normalized on the critical current of the referenced Josephson junctions $I_\text{c}$. It has to be noticed that, being the rf-SQUIDs the nonlinear core of the modeled JTWPA, the driving pump current $I_\text{p}$ flowing through the amplifier can exceed the critical current of the Josephson junctions $I_\text{c}$ (i.e., the pump current can be unevenly divided between the two branches of the rf-SQUIDs, maintaining the junctions in their superconductive state). The gain results negligible for small values of $I_\text{p}$, whereas an increase in the pumping current increases the intensity of coupling between tones and thus the gain of the amplifier, with a corresponding reduction of the bandwidth. The dashed lines in Fig. \ref{fig:Gain} represent the edges of this bandwidth, defined exploiting a gain threshold of 3 dB.

To evaluate the performance of an ideal phase-preserving linear amplifier it is possible to define its quantum efficiency spectrum as the ratio between the vacuum fluctuations in the input field and the fluctuations of the output field, this latter quantity normalized on the bimodal gain of the amplifier \cite{Renger2021}
\begin{equation}
    \eta(\omega)\equiv\frac{\braket{(\Delta\hat{a}^{\text{vac}}_{\omega,\text{in}})^2}
    }{\braket{(\Delta\hat{a}_{\omega,\text{out}})^2}/G_\text{b}(\omega)}=\frac{1/2\cdot G_\text{b}(\omega)}{\braket{(\Delta\hat{a}_{\omega,\text{out}})^2}}
\end{equation}
where $\braket{(\Delta\hat{a}_{\omega,\text{in (out)}})^2}=\braket{\hat{a}^2_{\omega,\text{in (out)}}}-\braket{\hat{a}_{\omega,\text{in (out)}}}^2$ is the variance of the input (output) annihilation operator with frequency $\omega$.
Another commonly exploited figure of merit to evaluate the performance of an amplifier is the noise figure spectrum $\mathcal{F}(\omega)$, defined as the ratio between the input signal to noise ratio (SNR) and the output SNR
\begin{equation}
    \mathcal{F}(\omega)\equiv\frac{\text{SNR}_{\text{in}}(\omega)}{\text{SNR}_{\text{out}}(\omega)}
\end{equation}
with
\begin{equation}
    \text{SNR}_{\text{in}}(\omega)\equiv\frac{\braket{\hat{n}_{\omega,\text{in}}}^2}{\braket{\left(\Delta\hat{n}_{\omega,\text{in}}\right)^2}}
\end{equation}
and
\begin{equation}
\label{eq:SNR}
    \text{SNR}_{\text{out}}(\omega)\equiv\frac{\braket{\hat{n}_{\omega,\text{out}}}^2-\braket{\hat{n}^{\text{vac}}_{\omega,\text{out}}}^2}{\braket{\left(\Delta\hat{n}_{\omega,\text{out}}\right)^2}}
\end{equation}
where $\braket{(\Delta\hat{n}_{\omega,\text{in (out)}})^2}=\braket{\hat{n}^2_{\omega,\text{in (out)}}}-\braket{\hat{n}_{\omega,\text{in (out)}}}^2$ is the variance of the input (output) photon number with frequency $\omega$. In the definition of the output SNR given in \eqref{eq:SNR} the contribution given by the added noise photons has been excluded from the amplified output field \cite{Shi2011}.

The definitions given so far hold for any bimodal input states. In the following subsections the quantum efficiency and the noise figure associated to two particular classes of them are presented.
\subsection{Bimodal Fock input states}
Considering a generic bimodal Fock state $\ket{\Psi_\text{F}}=\ket{n_{\omega}}_{\omega}\ket{n_{\omega'}}_{\omega'}$ as the input state of the amplifier, it can be easily demonstrated that the bimodal gain is
\begin{equation}
    G_{\text{b},\text{F}}(\omega)=|u(\omega)|^2+\frac{n_{\omega'}}{n_{\omega}}|v(\omega)|^2
\end{equation} 
and the variance of the output annihilation operator at $\omega$ frequency is $\braket{(\Delta\hat{a}_{\omega,\text{out}})^2}_{\text{F}}=1/2\cdot[(1+2n_{\omega})|u(\omega)|^2+(1+2n_{\omega'})|v(\omega)|^2]$. Exploiting \eqref{eq:UnitaryCondiction} and \eqref{eq:gain} the quantum efficiency results in
\begin{equation}
    \eta_\text{F}(\omega)=\frac{G(\omega)+(G(\omega)-1)\cdot(n_{\omega'}/n_{\omega})}{2G(\omega)[1+n_{\omega}+n_{\omega'}]-2n_{\omega'}-1}
\end{equation}
In the high gain limit (i.e., $G(\omega)\gg1$) this quantity tends to
\begin{equation}
    \eta_{\text{F}(\omega)}\rightarrow\frac{n_{\omega}+n_{\omega'}}{2n_{\omega}(1+n_{\omega}+n_{\omega'})}
\end{equation}
In this limit, supposing the input $\omega'$ mode in the vacuum state (i.e., $n_{\omega'}=0$) the quantum efficiency reduces to $\eta_\text{F}(\omega)=1/[2(1+n_{\omega})]$, at most equal to the standard quantum limit (SQL) $\eta_{\text{SQL}}(\omega)=1/2$. The introduction of an $\omega'$ field at the input port (i.e., $n_{\omega}\neq0$) induces an increase of the quantum efficiency of the amplifier, up to the limit $1/(2n_{\omega})$.

\begin{figure}
    \centering
    \includegraphics[width=87.5mm]{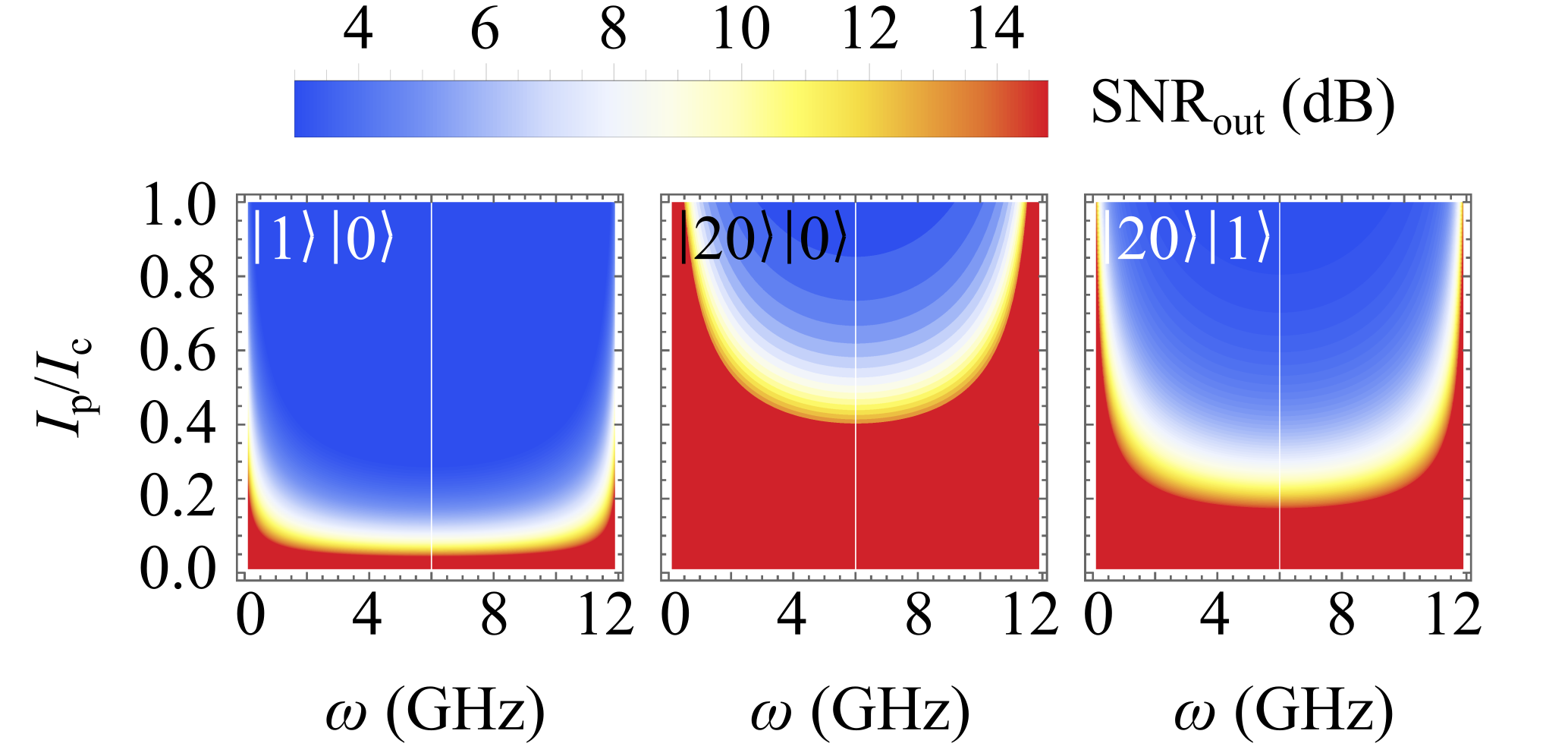}
    \caption{Signal to Noise Ratios of the output field at $\omega$ frequency $(\text{SNR}_{\text{out}}(\omega))$, expressed in dB, for three different bimodal Fock state inputs as a function of $I_\text{p}$ normalized on the critical current $I_\text{c}$ of the Josephson junctions composing the referenced device.}
    \label{fig:SNR Fock States}
\end{figure}

Regarding the noise figure spectrum, it can be derived that $\mathcal{F}_{\text{F}}(\omega)=\infty$, being $\text{SNR}_{\text{in}}(\omega)=\infty$, for any value of $n_{\omega}$ and $n_{\omega'}$. Fig.\;\ref{fig:SNR Fock States} presents the $\text{SNR}_{\text{out}}(\omega)$ for three different bimodal Fock input states as a function of $I_\text{p}/I_\text{c}$. For a given input state, across all the frequency spectrum, the $\text{SNR}_{\text{out}}(\omega)$ decreases with the increase of $I_\text{p}$ (thus with the raise of the gain of the amplifier, as shown in Fig.\;\ref{fig:Gain}). On the contrary, when $I_\text{p}$ tends to zero, the amplifier slightly perturbs the input state, and the $\text{SNR}_{\text{out}}(\omega)$ diverges, recovering the value of $\text{SNR}_{\text{in}}(\omega)$.
In accordance with the gain profile, a drastic reduction of the $\text{SNR}_{\text{out}}(\omega)$ occurs for $I_\text{p}$ values that increase with the increase of the distance of the considered mode $\omega$ from the center of the band of the amplifier.
Furthermore, for a fixed value of $I_\text{p}$, an increase of the occupancy of the input $\omega$ mode induces a raise of the $\text{SNR}_{\text{out}}(\omega)$, while an increase of the occupancy of the input $\omega'$ mode induces a reduction. Therefore, the drawback of the enhancement of the output power at $\omega$ frequency promoted by an $\omega'$ input field is the degradation of the $\text{SNR}_{\text{out}}(\omega)$.
\subsection{Bimodal Coherent input states}
In this subsection a bimodal coherent state $\ket{\Psi_\text{C}}=\ket{\alpha_\omega}_\omega\ket{\alpha_{\omega'}}_{\omega'}=D(\alpha_\omega)\ket{0}_\omega D(\alpha_{\omega'})\ket{0}_{\omega'}$, with $D$ the displacement operator, is considered as the input state of the amplifier. For this  input state the bimodal gain results to be
\begin{align}
    &G_\text{b,C}(\omega)=\nonumber\\
    &=|u(\omega)|^2+\frac{\alpha^2_{\omega'}}{\alpha^2_{\omega}}|v(\omega)|^2+i\frac{\alpha_{\omega'}}{\alpha_{\omega}}\left(v(\omega) u^{*}(\omega)+u(\omega)v^{*}(\omega))\right)
\end{align}
while the variance of the output annihilation operator is $\braket{(\Delta\hat{a}_{\omega,\text{out}})^2}_\text{C}=1/2\cdot(|u(\omega)|^2+|v(\omega)|^2)$.
\begin{figure}[b]
    \centering
    \includegraphics{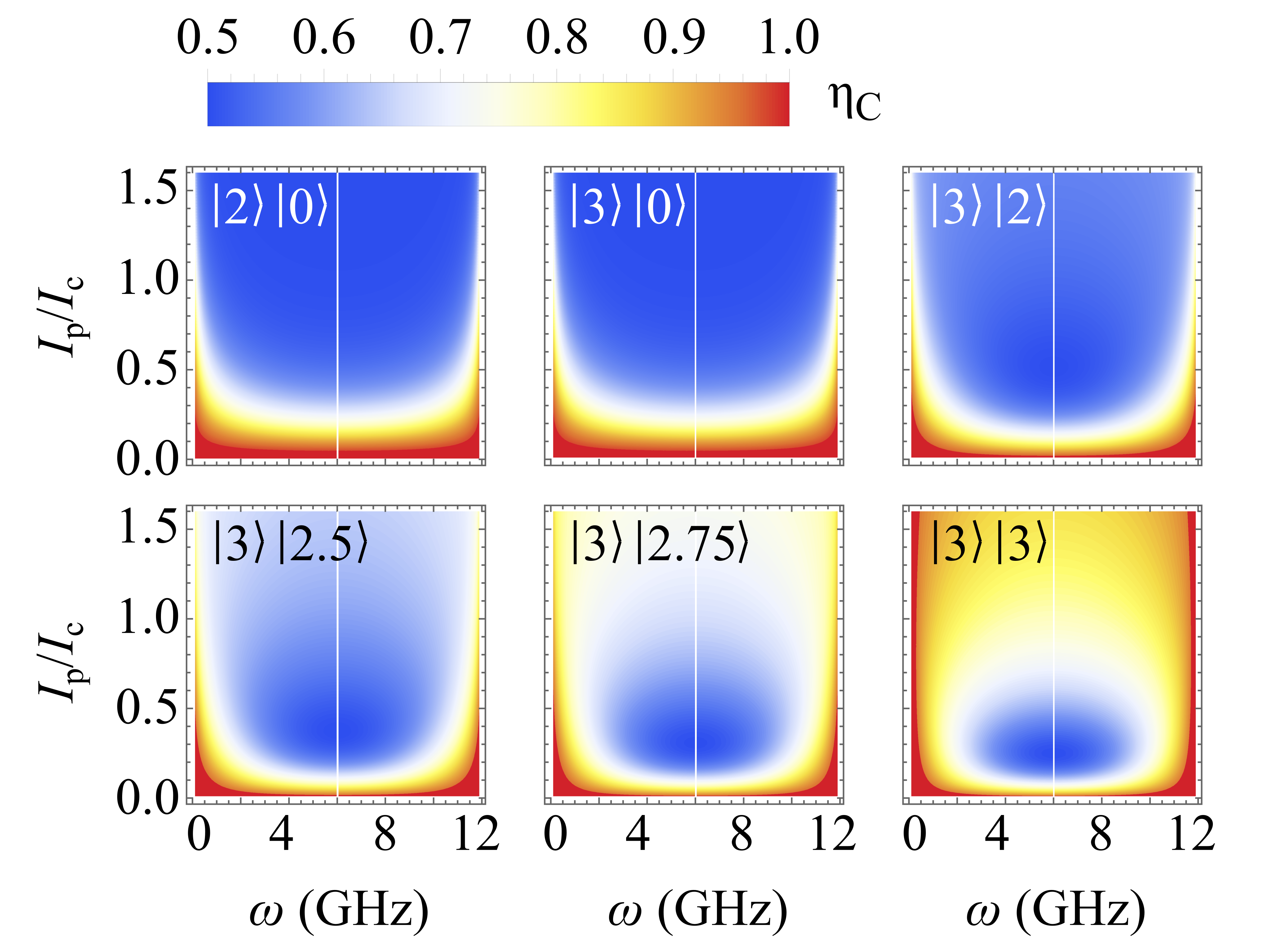}
    \caption{Quantum efficiency spectrum $(\eta_{C}(\omega))$ for six different bimodal coherent input states as a function of $I_\text{p}$ normalized on the critical current $I_{\text{c}}$ of the Josephson junctions composing the referenced device.}
    \label{fig:QE_Coherent}
\end{figure}
Fig. \ref{fig:QE_Coherent} presents the spectral distribution of the quantum efficiency for six different bimodal coherent input states as a function of $I_{\text{p}}/I_{\text{c}}$. The shape of the distributions recalls once again the one of the gain $G(\omega)$. It can be observed that, when the $\omega'$ mode of the bimodal coherent input state is in the vacuum state (i.e., $\alpha_{\omega'}=0$), the quantum efficiency in the high gain limit tends to the SQL $\eta_{\text{SQL}}=1/2$. Differently from what described in the previous subsection, this limit is exceeded when the average occupancy of the input $\omega'$ mode is non-zero. The former result clearly indicates that going beyond the SQL is non limited to a linear amplifier working in the phase-sensitive regime (i.e., when $\omega=\omega'$), but it can also be obtained in the phase-preserving regime, in presence of a bimodal coherent input state with a non-zero average occupancy of the $\omega'$ mode.
In a complementary way, the exceed of the SQL can also be shown evaluating the noise figure for different coherent input states. Being for these states $\text{SNR}_{\text{in}}(\omega)=|\alpha_{\omega}|^2$, the noise figure can be written as $\mathcal{F}_{\text{C}}(\omega)=|\alpha_{\omega}|^2/\text{SNR}_{\text{out}}(\omega)$.

\begin{figure}[t]
    \centering
    \includegraphics[width=87.5mm]{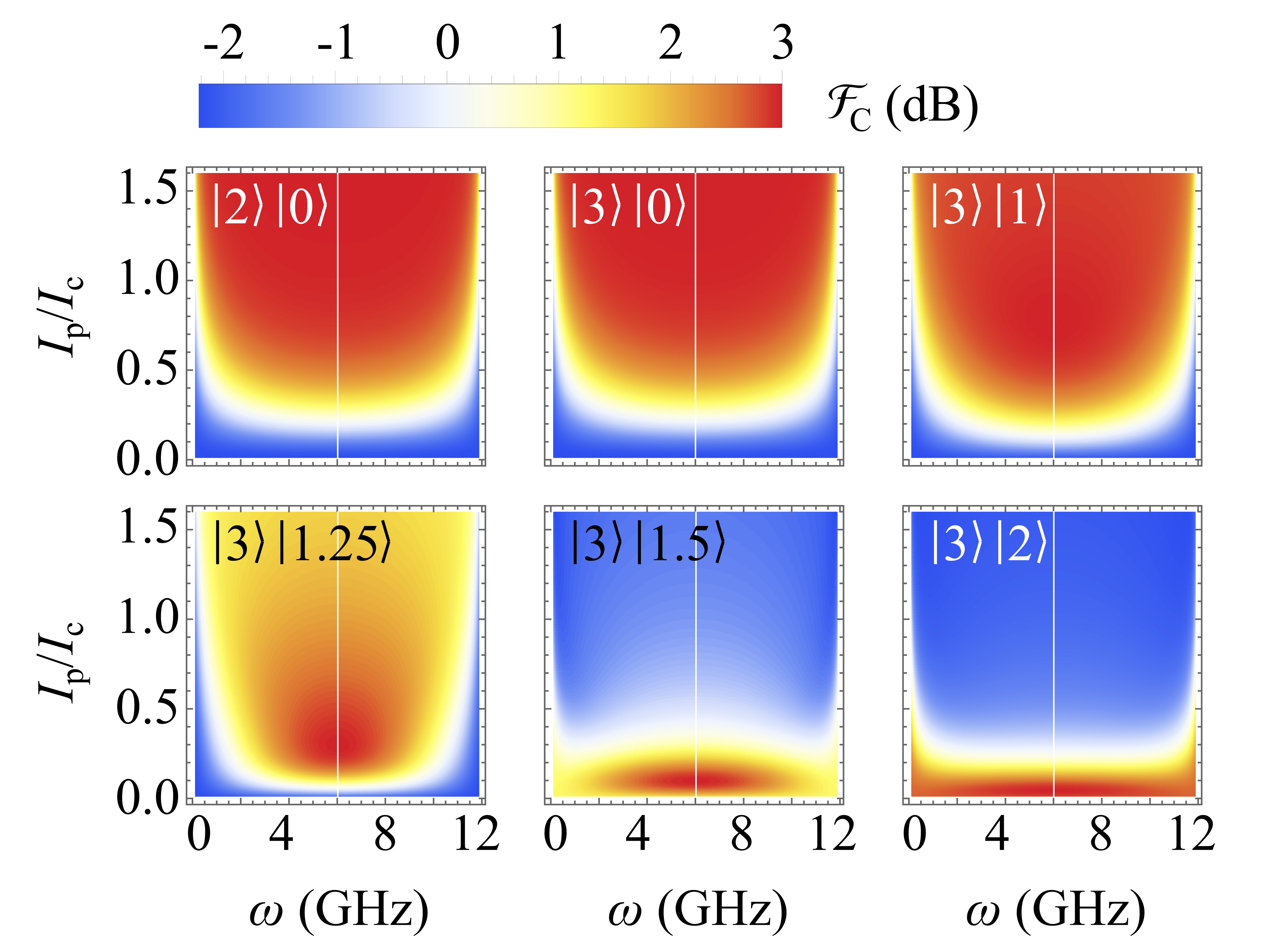}
    \caption{Noise figure $(\mathcal{F}_{\text{C}}(\omega))$, expressed in dB, for six different bimodal coherent states input as a function of $I_\text{p}$ normalized on the critical current $I_c$ of the Josephson junctions composing the referenced device.}
    \label{fig:NoiseFigureCoherent}
\end{figure}

Fig.\;\ref{fig:NoiseFigureCoherent} reports the spectral distribution of the noise figure for six different bimodal coherent input states as a function of $I_\text{p}/I_\text{c}$. Similarly to what previously discussed, for a fixed value of $I_\text{p}$, an increase in the average occupancy of the input $\omega$ mode (i.e., $\alpha_\omega$) induces an increase of the noise figure. In the high gain limit, when $\alpha_{\omega'}=0$, this quantity tends to the SQL $\mathcal{F}_{\text{SQL}}(\omega)=2$, whereas an increase in the average occupancy of the input $\omega'$ mode (i.e., $\alpha_{\omega'}$) allows to reach values beyond this limit.
\section{Effective Temperature}
\begin{figure}[t]
    \centering
    \includegraphics[width=87.5mm]{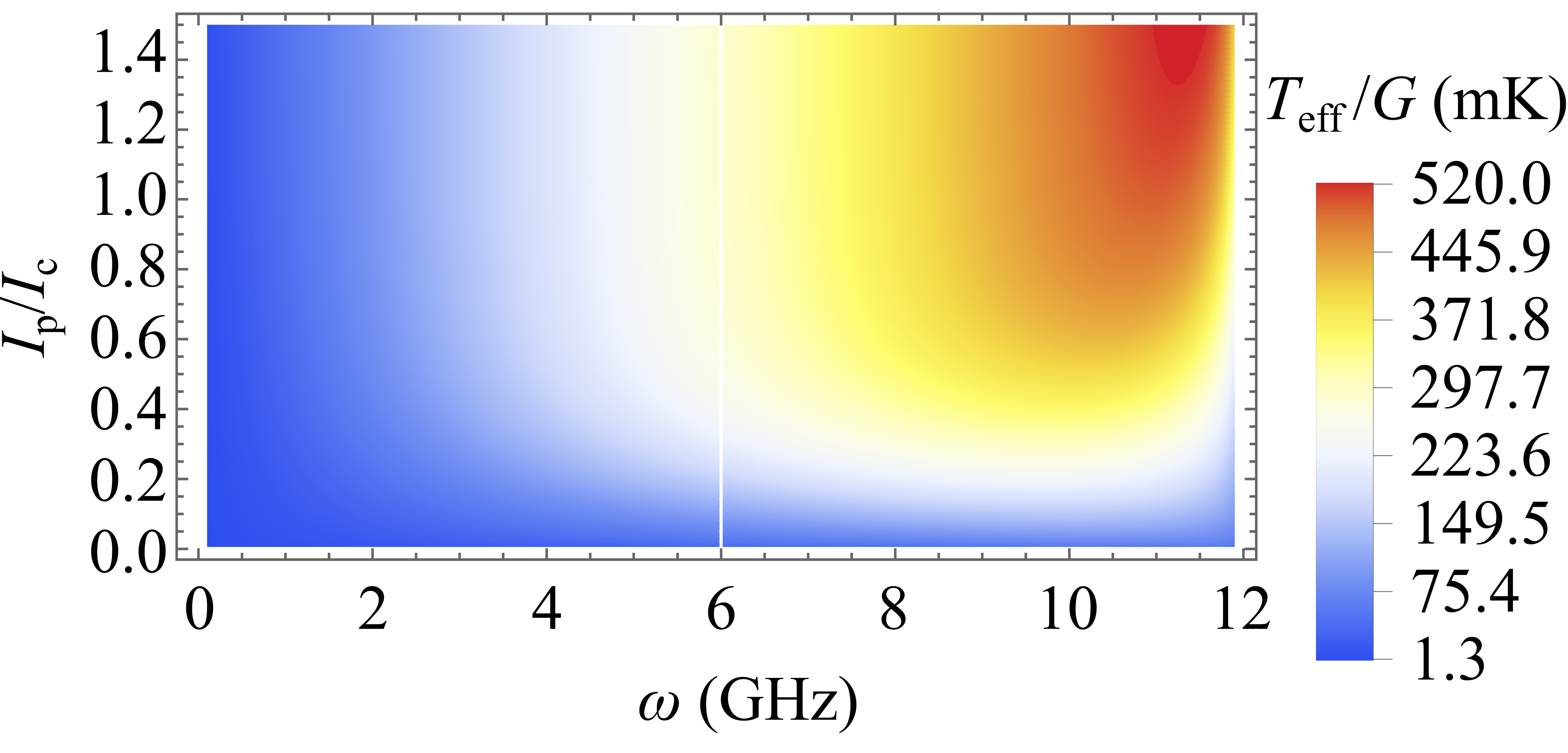}
    \caption{Spectral distribution of the normalized effective temperature $T_\text{eff}(\omega)/G(\omega)$,} as a function of $I_\text{p}$ normalized  on  the critical current $I_\text{c}$ of the Josephson junctions composing the referenced device.
    \label{fig:Teff}
\end{figure}

The quantification of the noise spectrum generated by an amplifier commonly requires the definition of its effective temperature $T_\text{eff}(\omega)$ as the temperature that a Bose-Einstein distribution should have to equal the output $\omega$ mode occupancy generated by a vacuum input state (see \eqref{eq:vacuum}) \cite{Clerk2010}
\begin{equation}
\label{eq:SignalModeOccupancy}
    \frac{1}{e^{\hbar\omega/k_\text{B}T_{\text{eff}}(\omega)}-1}=|v(\omega)|^2
\end{equation}
where $\hbar$ is the reduced Planck constant and $k_{\text{B}}$ is the Boltzmann constant. The expression for the effective temperature derives from \eqref{eq:SignalModeOccupancy} exploiting the unitary condition given in \eqref{eq:UnitaryCondiction} and the gain definition given in \eqref{eq:gain}
\begin{align}
    T_{\text{eff}}(\omega)&=\frac{\hbar\omega}{k_\text{B}}\left[\ln{\left(1+\frac{1}{|v(\omega)|^2}\right)}\right]^{-1}=\nonumber\\
    &=\frac{\hbar\omega}{k_\text{B}}\left[\ln{\left(1+\frac{1}{G(\omega)-1}\right)}\right]^{-1}
\end{align}
which implies that an ideal phase-preserving linear amplifier acting as a passive element (i.e., $G=1$) has an effective temperature equal to zero while, when acting as an active element (i.e., $G>1$), its effective temperature increases with the increase of the gain. Fig.\;\ref{fig:Teff} reports the numerical evaluation of the spectral distribution of the normalized effective temperature $T_\text{eff}(\omega)/G(\omega)$ for the referenced device as a function of $I_\text{p}/I_\text{c}$. Here it can be noticed that, in the high gain limit, the ratio $T_{\text{eff}}(\omega)/G(\omega)$ tends to the SQL $\hbar\omega/k_{\text{B}}$, corresponding to the sum of the equivalent temperature of a half of quantum of noise associated to the signal mode, plus the equivalent temperature of the minimal added noise that the idler mode has to inject in the $\omega$ mode in order to preserve the bosonic commutation relations at the output \cite{Peng2021}.
\section{Environment contribution to the Output Noise}
\begin{figure}[t]
    \centering
    \includegraphics[width=87.5mm]{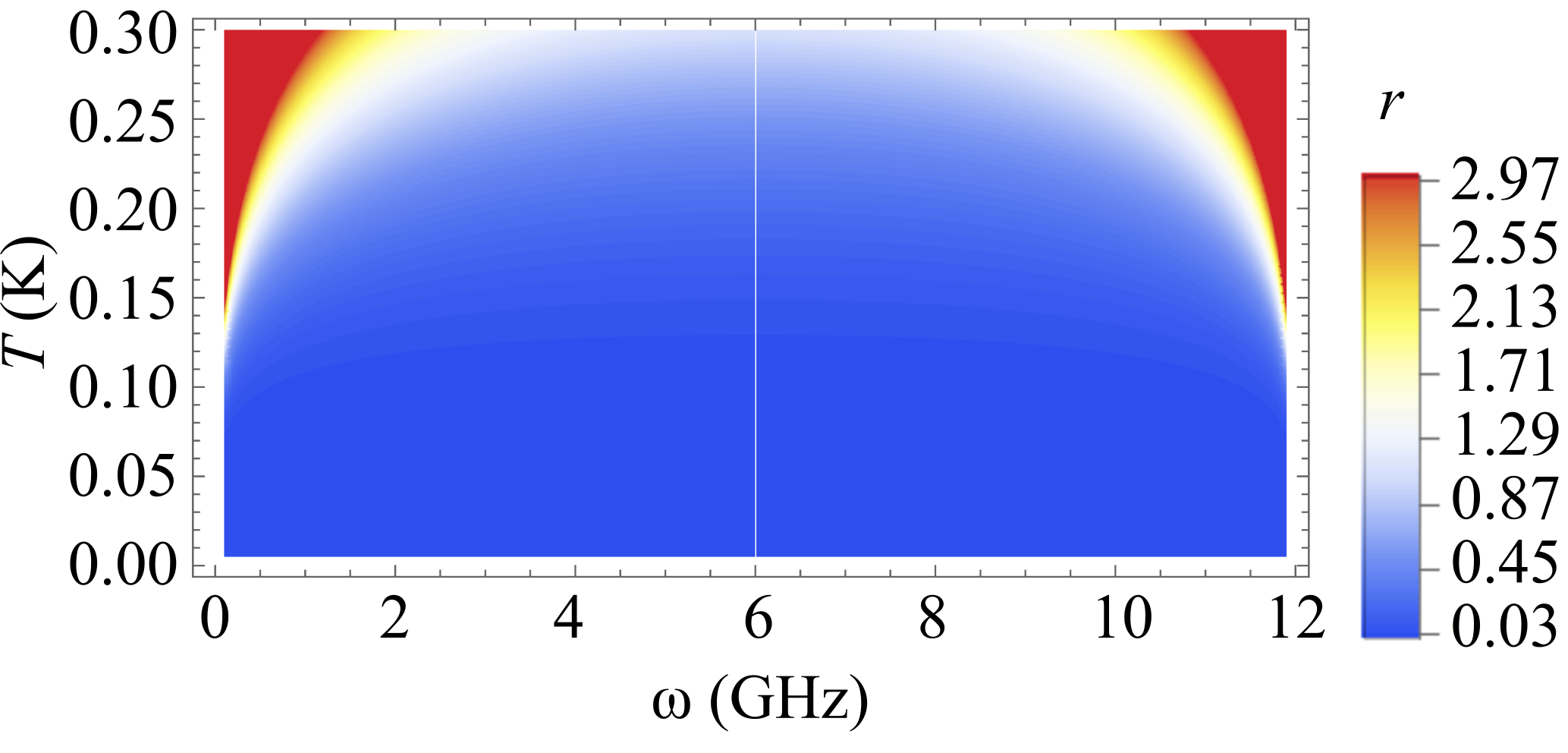}
    \caption{Spectral distribution of the adimensional ratio $(r(\omega))$ between the expectation value of the output photon number produced by the amplification of an input thermal-state and the added noise photons, as a function of the temperature $T$ of the environment. For this numerical computation the pump current was set to $I_\text{p}=I_\text{c}$.}
    \label{fig:Thermal_States}
\end{figure}

In this section the behaviour of the amplifier when inserted in a realistic measurement setup is taken into account. In particular the case in which the input field of the amplifier is a bimodal thermal-noise state generated by the surrounding environment at a given temperature $T$ is considered. The density operator for this state can be written as
\begin{equation}
    \rho_{\text{th}}=\sum_{n=0}^\infty\frac{e^{n
\beta\hbar\omega}}{1-e^{\beta\hbar\omega}}\ket{n}_{\omega}\bra{n}_{\omega}\sum_{m=0}^\infty\frac{e^{m\beta\hbar\omega'}}{1-e^{\beta\hbar\omega'}}\ket{m}_{\omega'}\bra{m}_{\omega'}
\end{equation}
where $\beta=1/k_{\text{B}}T$.
The expectation value for the photon number of the output field at frequency $\omega$ is obtained tracing over both the two modes Fock basis
\begin{align}
    \braket{\hat{n}
_{\omega,\text{out}}}_\text{th}&=\text{Tr}\left[{\rho_{\text{th}}\;\hat{n}_{\omega,\text{out}}}\right]=\nonumber\\
&=\sum_{j,k=0}^\infty\bra{j}_{\omega}\bra{k}_{\omega'}\cdot\rho_{\text{th}}\;\hat{n}_{\omega,\text{out}}\cdot\ket{j}_{\omega}\ket{k}_{\omega'}
\end{align}
To isolate the contribution to the output field given by the amplification of the thermal-noise and compare it with the contribution given by the added noise generated by the device itself, we define the ratio $r(\omega)=(\braket{\hat{n}_{\omega,\text{out}}}_\text{th}-\braket{\hat{n}^{\text{vac}}_{\omega,\text{out}}})/\braket{\hat{n}^{\text{vac}}_{\omega,\text{out}}}$. Fig.\;\ref{fig:Thermal_States} presents this quantity as a function of both the mode frequency and the temperature of the surrounding environment for the referenced device. 
It can be appreciated that in the low temperature regime the main contribution to the output noise, across the entire frequency spectrum, is given by the added noise ($r(\omega)<1$), while an increase of $T$ induces an increase of $r(\omega)$ as fast as the considered mode $\omega$ is far from the center of the band of the amplifier.
\section{Conclusion}
The behaviour of a generic phase-preserving linear amplifier having at the input port a bimodal signal composed by two uncorrelated fields at frequency $\omega$ and $\omega'=\omega_\text{p}-\omega$ was investigated. For this, a definition of the spectra of gain, quantum efficiency, noise figure, and effective temperature was given. Afterwards, the behaviour in terms of noise estimators of a rf-SQUID based JTWPA driven at different conditions and with different input states was simulated, as a case study of the theoretical framework. It was demonstrated that feeding the amplifier with bimodal Fock input states leads to a quantification for the noise estimators that is within the SQLs, while when exploiting bimodal coherent input states these limits can be exceeded.
The results represent a key model to predict the behaviour of rf-SQUID based JTWPA \cite{Zorin2016, Zorin2017} and other Josephson-based metamaterials \cite{Esposito2021} for quantum computing (via quantum limited amplification of complex signals) \cite{Macklin2015, Abdo2011, Aumentado2020} and for quantum information (via the generation of non-classical radiation \cite{Lloyd2008, Tan2008, Barzanjeh2015, Grimsmo2017, Fasolo2021}) in non-ideal cryogenic environment \cite{Krinner2019}.

\end{document}